\documentstyle[preprint,aps,prl]{revtex} 
\input psfig.sty

\def\tri{{{}^3{\rm H}}}
\def\hel{{{}^3{\rm He}}}
\def\qn{{\gamma}}
\begin{document}
\draft
\title{Neutron--${}^3{\rm H}$ and Proton--${}^3{\rm He}$ Zero Energy
Scattering} 
\author{ M. Viviani$^1$, S.~Rosati$^{1,2}$ and A.~Kievsky$^1$,}
\address{ $^1$Istituto Nazionale di Fisica Nucleare, Piazza Torricelli 2, 
56100 Pisa, Italy }
\address{ $^2$Dipartimento di Fisica, Universita' di Pisa, Piazza 
Torricelli 2, 56100 Pisa, Italy }
\date{\today}
\maketitle
\begin{abstract}
The Kohn variational principle and the (correlated)
Hyperspherical Harmonics  technique are applied to study the 
$n-\tri$ and $p-\hel$ scattering at zero energy. Predictions 
for the singlet and triplet scattering lengths are obtained for 
non--relativistic nuclear Hamiltonians including two-- and three--body 
potentials. The calculated  $n-\tri$ total cross section agrees well
with the measured value,  while some small discrepancy is found for
the coherent  scattering length. For the $p-\hel$ channel, the
calculated scattering lengths are in reasonable agreement with the values
extrapolated from the measurements made above 1  MeV. 
\end{abstract}

\narrowtext

In the last few years the scattering of nucleons by deuterons has been
the subject of a large  
number of investigations. This scattering problem is in 
fact a very useful tool for testing the accuracy of our
present knowledge of 
the nucleon--nucleon (NN) and three nucleon (3N) interactions. 
Noticeable progress has been achieved, but a number of relevant
disagreements between theoretical predictions and experimental results  remains
to be solved~\cite{GWHKG96,KRTV96}.  

It is therefore of interest to extend the above mentioned analysis 
to four nucleon scattering processes. In this case, an important goal for both 
theoretical and experimental analysis is to reach a 
precision comparable to that  achieved in the $N-d$ case. This is 
particularly challenging from the theoretical point of view, since the study 
of $A=4$ systems is noticeably more complicated than the $ A=3$ one.
Recently, accurate calculations of the alpha particle binding
energy $B_4$ have been achieved~\cite{PPCPW97,Varga97,KRV95}. 
It has been shown that, with NN+3N potential models fitting the
$\tri$ binding energy, no four--nucleon potential seems necessary to
reproduce the experimental value of $B_4$~\cite{PPCPW97}. Therefore, it is  
expected that NN and 3N interactions should be sufficient to describe
the four nucleon scattering processes too. Thus, discrepancies
between theory and experiment would be useful
to gain further information on the nuclear interaction. For example,
the polarization observables in the reaction $p-\tri$ are believed to
be very sensitive to the spin--orbit interactions~\cite{T65}. 
Moreover, four nucleon reactions play an important 
role also in Astrophysics and other subfields of physics.

In this letter, the problem
of $n-\tri$ and $p-\hel$ zero energy  scattering is studied.
The aim is to obtain accurate estimates  
of the corresponding  scattering observables by using
NN  and 3N realistic interactions.
The relevant quantities in $n-\tri$ zero--energy scattering are the
singlet $a_s$ and triplet $a_t$ scattering lengths. They  can be
obtained from the experimental values of the total cross section $\sigma_T$
and the coherent scattering length $a_c$,
\begin{equation} 
   \sigma_T= \pi (|a_s|^2+3|a_t|^2)\ , \quad 
    a_c={1\over4} a_s + {3\over4} a_t\ . \label{eq:def}
\end{equation}
The  $n-\tri$ cross section
has been accurately measured over a wide energy range
and the extrapolation to zero energy does not present any problems. The 
value obtained is $\sigma_T=1.70\pm 0.03$ b~\cite{Phill80}. The coherent
scattering  length has been measured by
neutron--interferometry techniques. The most recent values reported in
the literature 
have been obtained by the same group; they
are $a_c=3.82\pm0.07$ fm~\cite{Rauch81} and  $a_c=3.59\pm0.02$
fm~\cite{Rauch85}, the latter value being obtained with a
more advanced experimental arrangement.
Recently, the estimation of $a_c=3.607\pm0.017$ fm has been
obtained from $p-\hel$ data  by using an approximate
Coulomb--corrected R--matrix theory~\cite{Hale90}.

The corresponding quantities for $p-\hel$ scattering are more
difficult to evaluate. Approximate values have been determined 
from effective range extrapolations to zero energy of data taken above
1 MeV, and therefore suffer large uncertainties~\cite{Tegner,Knutson}.

From the theoretical point of view, the problem of the scattering of four 
nucleons has been considered for a  long time (see
ref.~\cite{Tilley92} and references cited therein). The most widely
used techniques are based on the Faddeev--Yakubovsky (FY)
equations~\cite{Tjon76,Fonseca79,Kharchenko76}
and the Kohn--Hulth\' en variational principles~\cite{Delves72}. In the latter
case, the Resonating Group Method (RGM) has been used to parametrize
the wave function (WF)~\cite{Hackenbroich70,Zahn81}, but also 
the expansion of the WF on a Hyperspherical Harmonic (HH) basis
has been investigated~\cite{Permyakov72}.  Calculations  using the
FY and HH techniques, which allow for the full  description of
the four body dynamics, were performed by using simple central or separable
potentials. Only recently, the  FY equations have been solved by
adopting  realistic NN potentials~\cite{Carbonell97}. 

In the present paper, the wave functions of the scattering states
are expanded in terms of the correlated Hyperspherical Harmonic
(CHH) basis~\cite{KRV93} and the  Kohn--Hulth\' en variational principles
are applied. Such a technique have been successfully used  in the study  
of the $N-d$ scattering below and above the deuteron breakup threshold. 
The present calculations follow exactly the same line followed in the
$N-d$ case described in ref.~\cite{KRV94}. Let us consider the
$p-\hel$ scattering; the case of $n-\tri$ scattering can be easily
obtained in the limit $e^2\rightarrow 0$, where $e$ is the unit charge (see
also ref.~\cite{Friar89}). The WF with total angular momentum $J$,
parity $\Pi$  and total isospin $T$, $T_z$ can be written as 
\begin{equation} 
  \Psi^{\qn}_{LS}=\Psi^\qn_C+\Phi^{\qn}_{LS} \ ,
   \label{eq:psitot} 
\end{equation}
where the index $\qn$  denotes hereafter the set of
quantum numbers $J,\Pi, T,T_z$.     
The first term $\Psi^\qn_C$ of eq.~(\ref{eq:psitot}) must be sufficiently
flexible to guarantee a detailed  description of the ``core'' of the
system, when all the particles are close to each other and the mutual
interaction is large; $\Psi^\qn_C$ goes to 
zero when the $p-\hel$ distance $r_{p}$  increases. 
It has been expanded in terms of CHH basis
functions, following the procedure  discussed in detail in
ref.~\cite{KRV95}.

The second term $\Phi^{\qn}_{LS}$
describes the asymptotic  configuration of the system, for large
$r_{p}$ values, where the nuclear $p-\hel$ interaction is
negligible. The quantum number $L$ is the relative orbital angular
momentum; $S$ is the spin obtained by coupling the   
spin $1/2$ of $\hel$ to the spin of the fourth nucleon. 
The angular momenta $L$ and $S$ are coupled to give the total angular
momentum $J$. 
In the present study the total isospin is $T= 1$.
The function $\Phi^{\qn}_{LS}$ must be the solution of the
two--particle Schroedinger equation appropriate for large $r_p$
values.  It is convenient to introduce the following surface functions
\begin{equation}
   \Omega^{(\lambda)}_{LS\qn}= \sum_{i=1}^4
       \left\{ Y_L({\hat r}_i) \left[ \Phi_{jk\ell}  \chi_i 
       \right]_{S} \right\}_{JJ_z}
       [\xi_{jk\ell} \xi_i]_{TT_z}
       {\cal R}^{(\lambda)}_L (r_i)
     \ ,\label{eq:surface1}
\end{equation}
where the product $\Phi_{jk\ell}\times\xi_{jk\ell}$ is the 
WF $\Psi_{{}^3{\rm He}}$ of the $\hel$ bound state (in the case of
$n-\tri$ scattering, it is the WF $\Psi_{{}^3{\rm H}}$ of the $\tri$
bound state). They are normalized to unity and are  antisymmetrical for
the exchange of any pair of particles $j$, $k$ and $\ell$. Both
$\Psi_{{}^3{\rm He}}$ and $\Psi_{{}^3{\rm H}}$ have been determined as
discussed in ref.~\cite{KRV94} by using the  CHH expansion for a
three--body system.  Within this scheme the WF and the binding energy
$B_3$ are determined with high accuracy. For example, the 
$B_3$ evaluated for  the different potential models considered in this
paper  agree within a few keV to the corresponding
results obtained by solving the Faddeev equations~\cite{CPFG89,WHG94}.

In eq.~(\ref{eq:surface1}), the spin (isospin) function  of
the unbound nucleon $i$ is denoted by $\chi_i$ ($\xi_i$). Moreover,
$r_i$ is the distance between nucleon $i$ and the center of mass of
$\hel$.  The functions ${\cal R}^{(\lambda)}_L (r_i)$ of
eq.~(\ref{eq:surface1})  can be taken to be  the regular
($\lambda\equiv R$) and irregular ($\lambda\equiv I$) radial solutions
of the two--body 
Schroedinger equation without nuclear interaction. They are analogous
to those used in $N-d$ scattering~\cite{KRV94}.

With the above definitions, the
asymptotic WF is written as 
\begin{equation}
   \Phi^{\qn}_{LS}= \Omega^{(R)}_{LS\qn} + \sum_{L'S'} 
         {}^\qn{\widetilde R}^{SS'}_{LL'}\Omega^{(I)}_{L'S'\qn}\ ,
         \label{eq:asympt} 
\end{equation}
where the matrix element ${}^\qn{\widetilde R}^{SS'}_{LL'}$ gives the
relative weight between the regular  and the irregular $L'S'$ components. 
These elements are the reactance matrix ($R$--matrix) elements, except
for some numerical factors~\cite{KRV94}. The eigenvalues of the
$R$--matrix are $\tan\delta_{LS}$, where $\delta_{LS}$ are the
eigenphase shifts of the ${}^{2S+1}L_{J\Pi TT_z}$ wave. 

The  convergence of
the expansion  of the internal part $\Psi^\qn_C$ is conveniently
studied by grouping the functions of the basis in  ``channels''
(a given channel contains CHH states with the same
angular--spin--isospin quantum numbers). It is very useful to
consider first the channels with orbital angular
momentum values as low as possible. One channel at a time is included in the
expansion of $\Psi^\qn_C$; the number
of the CHH functions belonging to that channel is increased until
convergence is reached. If the contribution of that particular channel
is found to be sizeable, the corresponding CHH functions  are retained
in the expansion; otherwise, they are rejected.
Then, others channels are added and the convergence
studied in terms of the total number of channels  $N_c$.
This procedure results to be effective since {\it i)\/}
the value of $N_c$  can be kept rather low, and {\it ii)\/} 
a small number of CHH 
functions is sufficient, except for few 
channels. In particular, for the states (S--wave,
$T=1$) considered here, the number of  channels included finally in
the wave functions is rather small   
($N_c\approx6\div8$). This is due mainly to the Pauli principle  which
prevents the overlap of the four nucleons. As a consequence,
the internal part is rather small and does not require a
large number of channels. 

The quantities to be determined in the WF~(\ref{eq:psitot}) are the
hyperradial  functions entering the HH expansion of
the internal part $\Psi^\qn_C$, and the matrix elements
$^\qn{\widetilde R}^{SS'}_{JJ'}$. For these,  the Kohn or the Hulth\' en
variational principles have been used. The Kohn variational principle
establishes that the following functionals 
\begin{equation}
  [{}^\qn {\widetilde R}^{SS'}_{LL'}]=
       {}^\qn{\widetilde R}^{SS'}_{LL'}-
       {M\over \sqrt6\hbar^2}
       \langle \Psi^\qn_{L'S'}|H-E|\Psi^\qn_{LS}\rangle\ ,
                     \label{eq:Kohn} 
\end{equation}
where ${}^\qn{\widetilde R}^{SS'}_{LL'}$ are the trial parameters
entering eq.~(\ref{eq:asympt}), must be stationary with respect to
variations of all the trial parameters of the WF. In
eq.~(\ref{eq:Kohn}, $E$ is the total (c.m.) energy and $M$ the nucleon
mass.

The form of the equations then derived and the procedure to solve
them is completely analogous to those of ref.~\cite{KRV94} and
is not repeated here.
With the Hulth\' en variational principle
the asymptotic function is written in the form
\begin{equation}
   \Phi^\qn_{LS}= \Omega^{(I)}_{LS\qn} + \sum_{L'S'} 
         {}^\qn{\widetilde U}^{SS'}_{LL'}\Omega^{(R)}_{L'S'\qn}\ ,
         \label{eq:asymptH} 
\end{equation}
where ${\widetilde U}={\widetilde R}^{-1}$. The Kohn and Hulth\' en
variational principles lead to essentially different
equations. Therefore, if the solutions in the two cases turn out to be
close to each other, we are quite confident that they are close to the
true solution.

The results for the singlet and triplet scattering lengths for $n-\tri$
scattering are given in table~\ref{tab:av14}, as a function of the
number of channels 
included in the WF. The potential adopted in this case is
the AV14 interaction~\cite{AV14}, so that a direct comparison with the
results obtained in ref.~\cite{Carbonell97} by solving the FY
equations can be made. From an inspection of the table, the
rapid convergence with $N_c$ is evident; this fact 
reflects that {\it i)\/} the
scattering lengths are mainly determined by the asymptotic part and
{\it ii)\/} the CHH expansion basis is very effective. Moreover, there
is a strict agreement between the converged values of the
scattering lengths obtained by means of the Kohn and the Hulth\' en
variational principles. Both estimates 
compare very well with the FY results of ref.~\cite{Carbonell97}, which is
a strong signal of the good accuracy of both
calculations. 

The calculated singlet and triplet $n-\tri$ scattering lengths
corresponding to different potentials models are plotted versus the
corresponding $\tri$ binding energy in fig.~\ref{fig:line}. 
The most recent experimental values~\cite{Rauch85,Hale90} of $a_s$ and
$a_t$ have also been reported. The models including
only NN forces are the
AV14~\cite{AV14}, AV8~\cite{AV8} and AV18~\cite{AV18} potentials.
Including 3N forces we have : the AV14+Urbana model VIII
(AV14UR)~\cite{URVIII}, 
AV18+Urbana model IX (AV18UR)~\cite{PPCPW97}, AV14+Brazil with 
$\Lambda=5.6 m_\pi$ (AV14BR1) and AV14+Brazil  with
$\Lambda=5.8 m_\pi$ (AV14BR2)~\cite{BR}. In the AV14UR and
AV18UR models, one the parameters of the 3N potentials was chosen so that to
reproduce the experimental $\tri$ binding energy value $B_3=8.48$
MeV. The AV14BR1 and AV14BR2 models have been chosen so as to
give slightly larger
$B_3$ values. It should be noted that all the results for the 
singlet (triplet) scattering length fall essentially on a straight
line. However, the experimental values extracted from the data do not
lie on the theoretical  
curves. This disagreement is related to a rather small discrepancy between
the calculated and measured coherent scattering length, as will be
shown below.

The calculated total cross section and coherent scattering length for
the AV14UR and AV18UR models are compared with the experimental 
values~\cite{Phill80,Rauch81,Rauch85,Hale90}
in table~\ref{tab:sigma}. These two potential models are chosen since
they well reproduce the experimental $B_3$ value, 
and  meaningful comparisons with the scattering data
extracted from experiments can be then
performed. From inspection of table~\ref{tab:sigma}, it can be
concluded that there is a 
satisfactory agreement between the calculated and the measured
value of $\sigma_T$.  The calculated coherent scattering lengths,
differ, however, by about 
3\% from the experimental values. This small discrepancy gives rise to
the large differences in the scattering 
lengths, when these are determined from the relations given in
eq.~(\ref{eq:def}). In fact, in the $a_s$, $a_t$ plane, the ellipse 
corresponding to the experimental values of the total cross section
$\sigma_T=1.7$ b and  the straight line corresponding
to the coherent scattering length  $a_c=3.7$ fm
are almost tangent. Therefore, a slight change in the $a_c$ value
produces a large variation of $a_s$ and $a_t$. This is also the
reason for the large uncertainty in the values of $a_s$ reported
in figure~\ref{fig:line}.

The $\hel$ binding energy $B_3(\hel)$ and the $p-\hel$ scattering
lengths as determined with the AV18 
and AV18UR models are presented in table~\ref{tab:phel}, together
with the  available experimental data~\cite{Tegner,Knutson}.
It should be remarked that, in contrast with the AV18UR model,  the
AV18 potential does not reproduce correctly the 
experimental value of $B_3(\hel)$. More in general, it has been
verified that the scattering 
length values show a scaling property analogous to that found in the
$n-\tri$ case. In table~\ref{tab:phel}, the available experimental
values have also been reported. However, it should be observed that {\it i)\/}
such experimental values have been extrapolated to zero energy
from measured data taken above $1$ MeV; {\it ii)\/} the quoted
``error bars'' include only statistical and not systematical
uncertainties~\cite{Knutson2}.  The $p-\hel$ experimental
scattering lengths therefore suffer large uncertainties, even somewhat
bigger than those reported in the table.
By inspection of the table it can be concluded that 
the agreement between the AV18UR predictions and the experimental
values is reasonably satisfactory and that it
would be very useful to have a more accurate experimental
determination of $a_s$ and $a_t$.
Finally, it should be noted that the $p-\hel$ scattering lengths are
larger than the corresponding values found in the $n-\tri$ case.  This
result is quite similar to that found in s--wave $N-d$ scattering
in the quartet spin state. 

In conclusion, accurate predictions of the $n-\tri$ and
$p-\hel$ zero energy scattering lengths with realistic hamiltonians
including NN and 3N potentials have been produced. The Kohn--Hulth\' en
variational principle and the correlated Hyperspherical Harmonics
technique were used to solve the four--body problem and to calculate
the quantities of interest. The singlet and triplet scattering lengths
for $n-\tri$ scattering were found to lie on straight lines when
plotted against the $\tri$ binding energy for a variety of potential
models. Our total cross section agrees well with the measured value, 
while some discrepancy is found in the comparison of the coherent 
scattering length values quoted in the literature. This is somewhat
surprising, since the corresponding quantity in $N-d$ scattering is
well reproduced by the theory~\cite{KRV94}, and the same was expected
for the four--nucleon case. 

Although low--energy
$p-\hel$ and $n-\tri$ experiments are difficult, we hope that
the present work might inspire further efforts in this area.

The authors wish to thank J. Carbonell and  L.D. Knutson for valuable
discussions and for providing their results prior to publication, and
I. Bombaci and L. Lovitch for the critical reading of the manuscript.

\begin{figure}[p]
\caption{Singlet (full symbols) and triplet (open symbols) 
scattering lengths plotted against the $\tri$ binding energy.
Circles labelled by a, b, c, d, e, f correspond to the AV18, AV14,
AV8, AV18UR, AV14BR1 and AV14BR2 models, respectively. The AV14UR
and AV18UR model predictions are almost coincident.
The squares (triangles) are the experimental values  of
ref.~\protect\cite{Rauch85} (ref.~\protect\cite{Hale90}).
The straight lines are linear fits of the theoretical results.}
\label{fig:line}
\end{figure}

\begin{table}
\begin{tabular}{ccdcd}
Method & $N_c$ & $a_s$ & $N_c$ & $a_t$ \\
\hline
K & 0 & 4.38 & 0 & 3.87 \\
K & 1 & 4.33 & 2 & 3.82 \\
K & 3 & 4.33 & 4 & 3.82 \\
K & 4 & 4.32 & 6 & 3.80 \\
K & 6 & 4.32 & 8 & 3.80 \\
\hline
H & 6 & 4.32 & 8 & 3.80 \\
\hline
FY & & 4.31 & & 3.79 \\
\end{tabular}
\caption{Singlet $a_s$ and triplet $a_t$ S-wave scattering lengths
(fm) for $n-\tri$ zero energy scattering calculated with the AV14
potential and the Kohn (rows labelled K) or
Hulth\' en (row labelled H) variational methods. $N_c$ is the number of
channels included in the CHH expansion of the wave functions (the case
$N_c=0$ corresponds to including in the WF only the asymptotic terms).
The last row reports the results obtained in
ref.~\protect\cite{Carbonell97} by solving the FY equations.} 
\label{tab:av14}
\end{table}

\begin{table}
\begin{tabular}{cdd}
Model & $\sigma_T$ &  $a_c$ \\
\hline
AV14UR & 1.74 & 3.71 \\
AV18UR & 1.73 & 3.71 \\
\hline
Expt.  & 1.70$\pm$0.03~\protect\cite{Phill80}  
        & 3.82$\pm$0.07~\protect\cite{Rauch81} \\
      &  & 3.59$\pm$0.02~\protect\cite{Rauch85} \\
      &  & 3.607$\pm$0.017~\protect\cite{Hale90} \\
\end{tabular}
\caption{Total cross section $\sigma_T$ (b) and coherent  scattering
length (fm) for $n-\tri$ zero energy scattering calculated with the AV14UR
and AV18UR potential models. The last rows report the
experimental values.}
\label{tab:sigma}
\end{table}

\begin{table}
\begin{tabular}{cddd}
Model & $B_3$ & $a_s$  &  $a_t$ \\
\hline
AV18   & 6.93 & 12.9 & 10.0 \\
AV18UR & 7.74 & 11.5 & 9.13 \\
\hline
Expt.  & 7.72 & 10.8$\pm$2.6~\protect\cite{Knutson} 
              & 8.1$\pm$0.5~\protect\cite{Knutson} \\
       &   &  & 10.2$\pm$1.5~\protect\cite{Tegner} \\
\end{tabular}
\caption{$\hel$ binding energy $B_3$ (MeV) and singlet $a_s$ and triplet
$a_t$ s-wave scattering lengths (fm)
for $p-\hel$  scattering calculated with the AV18
and AV18UR potential models. The last rows report the
experimental values.}
\label{tab:phel}
\end{table}

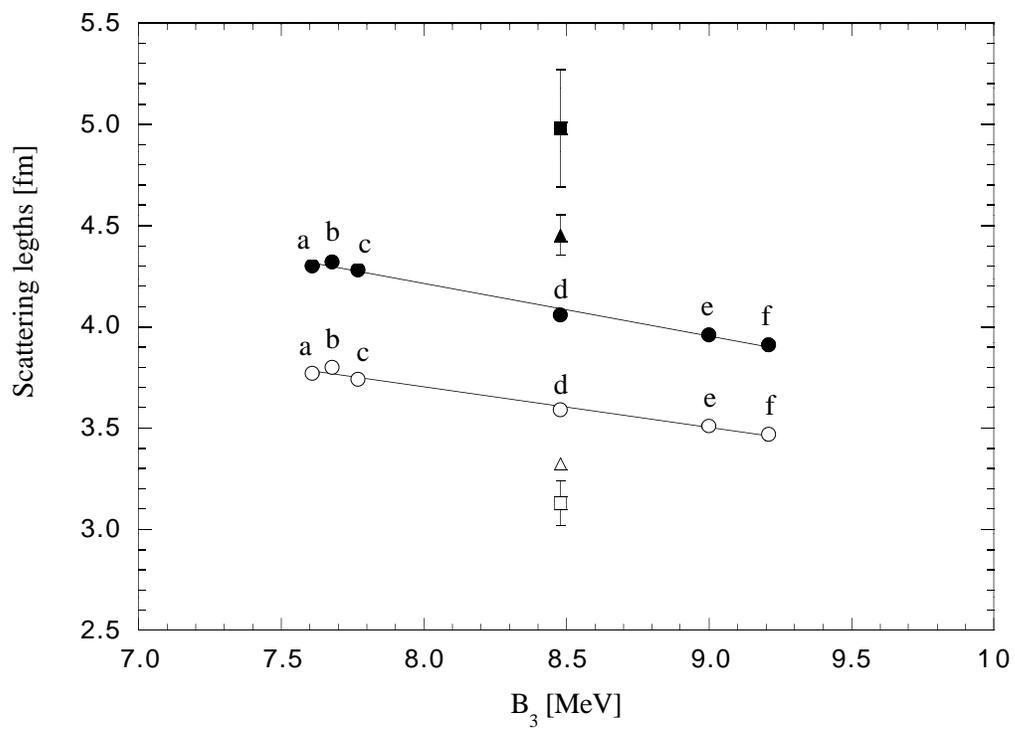
\end{document}